\documentclass[%
 reprint,
superscriptaddress,
 amsmath,amssymb,
 aps,
]{revtex4-2}

\usepackage{graphicx}
\usepackage{dcolumn}
\usepackage{bm}
\usepackage{lineno}
\usepackage{amsfonts}
\usepackage{makeidx}
\usepackage{colortbl}
\usepackage{multirow}
\usepackage{lmodern}
\usepackage{afterpage}
\usepackage{array}
\usepackage{physics}
\usepackage{upgreek}
\usepackage{xcolor}
\usepackage{xr} 

\usepackage[explicit]{titlesec}
\titlespacing\section{0pt}{20pt plus 4pt minus 2pt}{6pt plus 2pt minus 2pt}
\titlespacing\subsection{0pt}{20pt plus 4pt minus 2pt}{6pt plus 2pt minus 2pt}
\titlespacing\subsubsection{0pt}{20pt plus 4pt minus 2pt}{6pt plus 2pt minus 2pt}

\usepackage{caption}
\usepackage{cuted} 

\newcommand{\ee}{\operatorname{e}}

\begin{document}

\preprint{APS/123-QED}

\title{Broadband biphoton source for quantum optical coherence tomography \\ based on a Michelson interferometer}

\author{Konstantin Katamadze}
\author{Anna Romanova}%
 \email{romanova.phys@gmail.com}
 \author{Denis Chupakhin}%
 \author{Alexander Pashchenko}%
 \author{Sergei Kulik}%
\affiliation{%
Quantum Technology Centre, Faculty of Physics, Lomonosov Moscow State University, Moscow, Russia
}%


\begin{abstract}
Broadband correlated photon pairs (biphotons) are valuable in quantum metrology, but current generation methods either involve complex nonlinear structures or lack sufficient bandwidth and brightness. In this work, we theoretically describe and experimentally demonstrate a novel technique for generation of a bright collinear biphoton field with a broad spectrum, achieved by using a tightly focused pump in a bulk nonlinear crystal. As the most straightforward application of the source, we employ Michelson interferometer-based quantum optical coherence tomography (QOCT). Utilizing the source enables the demonstration of record resolution and dispersion cancellation for this QOCT scheme.
\end{abstract}

\maketitle


\section{\label{sec:level1} Introduction}

Correlated photon pairs (biphotons) obtained through  spontaneous parametric down-conversion (SPDC)~\cite{Klyshko1988a} are one of the primary tools in quantum optics. 
The key role is played by time correlations between the photons of a pair, which become stronger the broader the spectrum of the biphoton field. 
A comprehensive review~\cite{Katamadze2022} details various implementations of broadband biphotons. 
Some are based on spatially homogeneous nonlinear media, where bandwidth is limited by phase matching and dispersion relations~\cite{Nasr2005, Carrasco2006a, peer_design_2006, hendrychBroadeningBandwidth2009, Katamadze2013, katamadzeBroadbandBiphotonsSingle2015, okothMicroscaleGenerationEntangled2019a, Javid2021}. 
Others use inhomogeneous media, where phase matching conditions vary, allowing broader bandwidth limited by spatial modulation range only~\cite{Carrasco2004a, nasrUltrabroadbandBiphotonsGenerated2008a, Kalashnikov2009, Katamadze2011a, Katamadze2011b, Okano2012, cao_efficient_2021, cao_non-collinear_2023}. 
Achieving a biphoton field with a bandwidth up to 150-200 THz is feasible with these techniques. 
However, achieving higher values is challenging because, in all the mentioned approaches (with the exception of~\cite{katamadzeBroadbandBiphotonsSingle2015}), spectral broadening is inherently linked to a decrease in spectral intensity. 
In the best-case scenario, integral intensity may be conserved, but this is typically divergent from real experimental conditions.

Broadband biphotons find applications in diverse quantum metrological problems. 
Examples include quantum clock synchronization~\cite{Giovannetti2001a,Giovannetti2001,Valencia2004,Quan2019,Quan2020}, two-photon microscopy~\cite{Giovannetti2004,Dowling2008,Varnavski2020}, two-photon spectroscopy~\cite{salehEntangledPhotonVirtualState1998,Dayan2005,Schlawin2018}. 
Another notable and straightforward application is quantum optical coherence tomography (QOCT)~\cite{abouraddy_quantum-optical_2002}, which extends the capabilities of classical optical coherence tomography (OCT)~\cite{huang_optical_1991}. 
Early OCT systems employed a broadband light source and a Michelson interferometer (MI), with one arm serving as a scanning reference and the other containing the test sample. 
Despite advancements in OCT's accuracy through intricate experimental designs and sophisticated data analysis, challenges persist, particularly in addressing limited scanning depth and resolution for highly dispersive samples.

A quantum version of OCT utilizes entangled photon pairs and the Hong-Ou-Mandel effect. 
This method, being dispersion-tolerant, offers double the resolution, reaching record values~\cite{okano54MmResolution2015}. 
The initial hurdle was the requirement for a bright broadband biphoton source, typically produced by non-collinear or type-II SPDC. 
However,  further research showed that QOCT could also be achieved using a brighter, broader, and more easily adjustable collinear type-I SPDC source, coupled with an MI~\cite{odateTwophotonQuantumInterference2005, lopez-mago_coherence_2012, lopez-mago_quantum-optical_2012, yoshizawa_telecom-band_2014}.

In this work, we experimentally study for the first time a novel biphoton broadening technique based on tightly focusing the pump and target modes~\cite{katamadze_broadband_2016}, and apply it to MI-based QOCT.
Ideally, this broadening approach preserves spectral intensity, leading to an increase in integral intensity with a broader spectral bandwidth.

The article unfolds as follows. 
In Section~\ref{sec:source}, we provide a theoretical description of the source's operation and detail its experimental implementation. Section~\ref{sec:interferometer} delves into the theoretical and laboratory aspects of the interferometer designed for QOCT. 
Finally, in Section~\ref{sec:results}, we present and discuss the experimental results of QOCT experiments.

\section{\label{sec:source}Bright source of broadband biphoton field}

\subsection{Theory}
\begin{figure}[h]
\centering
\includegraphics[width=\linewidth]{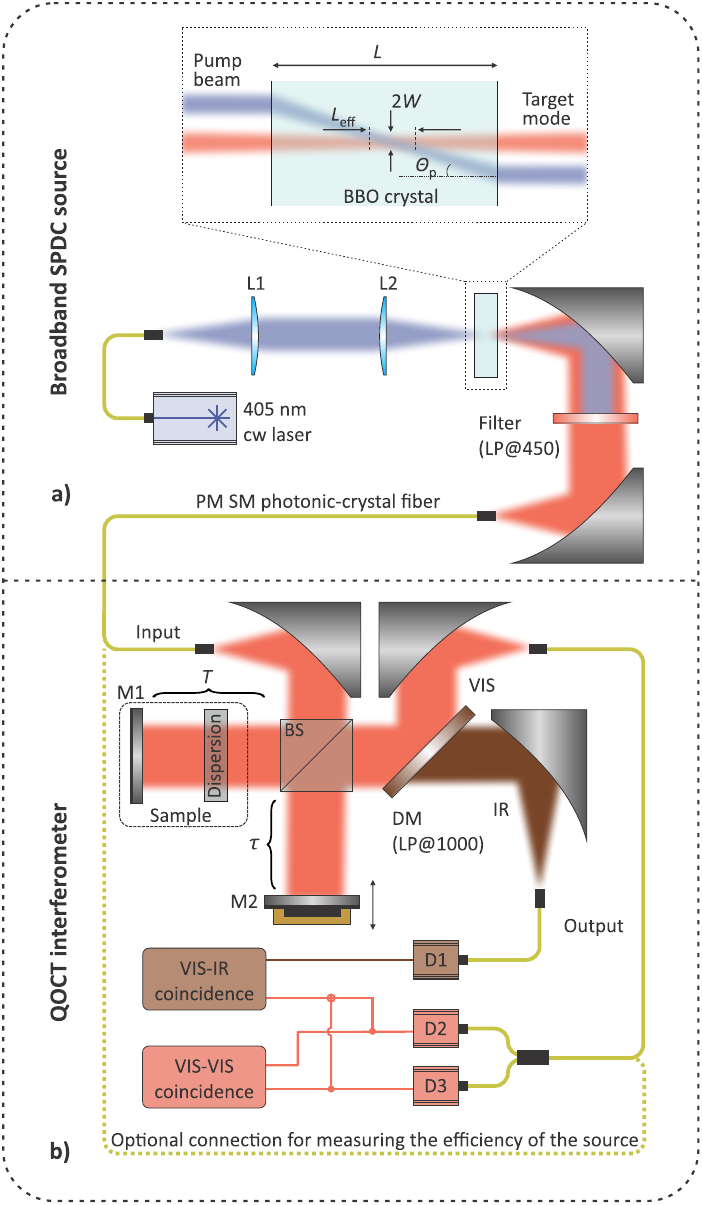}
\caption{Experimental setup. a) Broadband biphoton source:
$L$, crystal length; $W$, modes waist; $\Theta_\text{p}$, pump walk-off angle;
L1, L2, lenses. b) QOCT Michelson interferometer: BS, beam splitter; DM, long-pass dichroic mirror splitting biphotons at 1000 nm; M1, reference mirror; M2, sample mirror, standing on the piezotranslator, which is attached to a mechanical translator; D1, single-photon detector (SPD), based on InGaAs avalanche
photodiode (APD), MPD PDM–IR; D2, D3, SPDs, based on Si
APDs, Laser Components COUNT NIR.
}
\label{fig:setup}
\end{figure}

A comprehensive theoretical description of our biphoton source can be found in~\cite{katamadze_broadband_2016}. Similar to the listed techniques~\cite{Nasr2005,Carrasco2006a,peer_design_2006,hendrychBroadeningBandwidth2009,Katamadze2013,katamadzeBroadbandBiphotonsSingle2015,okothMicroscaleGenerationEntangled2019a,Javid2021,Carrasco2004a,nasrUltrabroadbandBiphotonsGenerated2008a,Kalashnikov2009,Katamadze2011a,Katamadze2011b,Okano2012,cao_efficient_2021,cao_non-collinear_2023}, we employ SPDC process as a biphoton source. SPDC involves the decay of pump (p) photons into pairs of signal~(s) and idler~(i) photons, satisfying energy conservation and phase-matching conditions:
\begin{align}
    \omega_\text{s}+\omega_\text{i}-\omega_\text{p}&=0,\\
    \vec{k}_\text{s}+\vec{k}_\text{i}-\vec{k}_\text{p}&=\vec{\Delta}_k,
\end{align}
where $\omega_\text{p,s,i}$ are frequencies, $k_\text{p,s,i}$ are wave vectors, and $\vec{\Delta}_k$ is a phase mismatch, typically $\lesssim$ the inverse nonlinear media length $1/L$.

Since our biphoton source interfaces with an interferometer, our focus is on the photon pair rate in a single spatial target mode, filterable by a single-mode fiber. 
We assume a Gaussian profile for the pump, signal, and idler modes propagating along the $z$-axis (top inset in Fig.~\ref{fig:setup}a):
\begin{equation}\label{eq:modes}
\begin{gathered}
   \vec{E}_j(\vec{r}, t)=\frac{1}{2}\left[\vec{E_j} g_j(\vec{r}) \ee^{-i \omega_j t}+\text { c.c. }\right], \\ \text{where }
\quad g_j(\vec{r})=\ee^{i k_j z} \ee^{-\frac{x^2+(y-\Theta_j z)^2}{W^2_j}}, \quad j=\text{p, s, i}. 
\end{gathered}
\end{equation}
Here, $\vec{E_j}$ is a field amplitude, $\Theta_j$ is a walk-off angle, and $W_j$ is a beam waist. 
According to~\cite{ling_absolute_2008}, the photon pair rate in the selected signal and idler modes can be calculated using a mode overlap integral:
\begin{multline}\label{eq:overlap}
R=\frac{8 d^2 P}{\pi^6 W^2_\text{p} W^2_\text{s} W^2_\text{i} c^3} \frac{\omega_\text{s} \omega_\text{i}}{n_\text{p} n_\text{s} n_\text{i}} \int_{-\infty}^{+\infty} \text{d} \omega_\text{s}\\
\times\left|\int_{-\infty}^{+\infty} \text{d} x \text{d} y \int_{-L / 2}^{L / 2} \text{d} z g_\text{p}(\vec{r}) g_\text{s}^*(\vec{r}) g_\text{i}^*(\vec{r})\right|^2,
\end{multline}
where $d$ is an effective second-order nonlinear susceptibility, $n_\text{p}, \ n_\text{s},\  n_\text{i}$ are refractive indices for the pump, signal, and idler modes, $c$ is the speed of light, and $P=c \epsilon_0 n_\text{p} \pi W_\text{p}^2\left(\left|E_\text{p} \right| / 2\right)^2$ is the pump power.

Considering a collinear e-oo type-I phase matching, i.e., $\Theta_\text{s}=\Theta_\text{i}=0$, and assuming equal waists for simplicity ($W_\text{p}=W_\text{s}=W_\text{i}=W$), we focus on a low-waist case where an effective overlap length $L_\text{eff} = W/\Theta_\text{p}\ll L$. 
Thus, in \eqref{eq:overlap}, we can integrate over $z$ from $-\infty$ to $+\infty$. After evaluating the integral, we obtain:
\begin{equation}\label{eq:2A_integral}
R=\underbrace{\frac{4 d^2 P \omega_\text{s} \omega_\text{i}}{3 \pi^3 c^3 \epsilon_0 n_\text{p} n_\text{s} n_\text{i} \Theta_{\text p}^2}}_{S_0} \underbrace{\int \ee^{ -3\left(\Delta_k \left(\omega_\text{s}\right) L_\text{eff} / 2\right)^2} \text{d} \omega_\text{s}}_{B_\omega}.
\end{equation}
Here, the factor $S_0$ represents the spectral coincidence rate, which is independent of the beam waist $W$ and crystal length $L$. The factor $B_\omega$ defines the integral spectral bandwidth and depends on the effective interaction length $L_{\text{eff}}$ instead of $L$. As a result, both the total rate $R$ and the bandwidth $B_\omega$ are proportional to $1/\sqrt{W}$.

In \eqref{eq:modes}, we do not account for beam divergence, assuming a Rayleigh length $z_R \gg L_\text{eff}$, which is equivalent to the inequality $W \gg \lambda_\text{p} / (\pi \Theta_\text{p})$. For the BBO crystal pumped at $\lambda_\text{p} = 405$~nm with collinear degenerate type-I phase matching, as studied in~\cite{katamadze_broadband_2016} and implemented in our experiment, the walk-off angle is $\Theta_\text{p} = 3.9^\circ$, and the waist should satisfy the inequality $W \gg 2~\upmu \rm m$.

Note that the spectral shape defined by the function $\ee^{ -3\left(\Delta_k \left(\omega_\text{s}\right) L_\text{eff} / 2\right)^2}$ is almost Gaussian and can be tailored by the shaping of the pump and target modes. This technique can be used to increase the SPDC heralding efficiency and single-photon spectral purity instead of employing complicated and non-adaptive non-linear domain engineering~\cite{branczyk_engineered_2011,bendixon_spectral_2013,tambasco_domain_2016}. 

For the case described in~\cite{katamadze_broadband_2016} and implemented in our experiment, the spectral coincidence efficiency is $S_0/P = 125\ \frac{\text{cps}}{\text{THz}\times\text{mW}}$, and the total spectral bandwidth is given by $B_\omega = \frac{\kappa}{\sqrt{W}}$, where $\kappa = 2\pi \times 298\ \text{THz} \times \sqrt{\upmu \rm m}$.

Therefore, the primary idea is to reduce the interaction length $L_\text{eff}$ and compensate for a small rate through high focusing of the pump and target modes. 
A similar approach was demonstrated in~\cite{okothMicroscaleGenerationEntangled2019a}, where authors used extremely short nonlinear media, achieving a significant biphoton rate due to high focusing. 
Conversely, another biphoton broadening technique using pump focusing~\cite{Carrasco2006a} is based on a quite different approach and cannot provide as extensive broadening as our technique.

\subsection{Implementation}

The schematic representation of the source is presented in Fig.~\ref{fig:setup}a. To achieve dispersion cancellation, as detailed in \cite{okanoDispersionCancellationHighresolution2013, Supplement}, we employed a narrowband pump laser (Toptica DLC DL pro HP 405) with a wavelength of 405~nm and a bandwidth of approximately 100~kHz. 
The laser radiation passed through a single-mode fiber and was then tightly focused onto a 1~mm-thick nonlinear crystal (BBO type-I). 
The resulting collinear degenerate biphoton field was collimated using a parabolic mirror and passed through a pump-blocking 450~nm longpass filter.
Subsequently, another parabolic mirror was utilized to couple the field into a single-mode polarization-maintaining photonic crystal fiber (NKT Photonics LMA-PM-10), chosen for its ability to guide and preserve the broadband spectrum. 
The biphoton field was then directed into the Michelson interferometer, described in the next section.

The alignment of the pump and target beams was meticulously optimized to achieve the maximum coincidence count rate, measured directly after the output of the fiber, reaching a final value of 3~kcps.
The pump power, measured in front of the crystal, was 8~mW. 
Post-experiment, the crystal was removed, and the pump waist in free space was measured as $W_p=5.7~\upmu \rm m$. This value is close to the critical threshold of $2~\upmu \rm m$, beyond which beam divergence should be accounted for. This could be a potential source of mismatch between the theoretical and experimental results.

\section{Interferometer}\label{sec:interferometer}

\subsection{Theory}

A comprehensive theoretical description of biphoton field interference in a Michelson interferometer (MI) can be found in~\cite{lopez-mago_coherence_2012, lopez-mago_quantum-optical_2012, Supplement}. Here, we will provide the main basis and results. 
Consider an input biphoton field in a single spatial mode with the following spectral quantum state:
\begin{equation}\label{Michelson1.Psiin}
    \left| {{\Psi }_{\text{in}}} \right\rangle =
    \frac{1}{\sqrt{2}}
    \iint{\text{d}{{\omega }_\text{s}}\text{d}{{\omega }_\text{i}}}f\left( {{\omega }_\text{s}},{{\omega }_\text{i}} \right)\hat{a}_{\text{in}}^{\dagger }\left( {{\omega }_\text{s}} \right)\hat{a}_{\text{in}}^{\dagger }\left( {{\omega }_\text{i}} \right)\left| \text{vac} \right\rangle. 
\end{equation}
where $f(\omega_\text{s},\omega_\text{i})$ is a spectral amplitude, approximated as a double-Gaussian function with the two-photon spectral distribution~\cite{Mikhailova2008, Fedorov2009} (Fig.~\ref{fig:Two_Omega}):
\begin{equation}\label{degenerate}
  {{\left| f({{\omega }_\text{s}},{{\omega }_\text{i}}) \right|}^{2}}=\frac{1}{\pi \delta \Delta }\underbrace{\ee^{-\frac{{{\left( \omega_\text{i}-\omega_\text{s} \right)}^{2}}}{2{{\Delta }^{2}}} }}_{\text{Phasematching}}\underbrace{\ee^ { -\frac{{\left(\omega_\text{i}+\omega_\text{s}-2\omega_0 \right)}^2}{2{{\delta }^2}}}}_{\text{Pump}}.
\end{equation}
Here, $\omega_0\equiv\omega_\text{p}/2$, $\delta$ is the standard deviation (StD) of the pump spectral distribution, and $\Delta=B_\omega/\sqrt{2\pi}$ is related to the phase matching bandwidth.

Now, let this field pass through an MI with a delay $T$ in a fixed arm and delay $\tau$ in a scanning arm (see Fig.~\ref{fig:setup}b and Fig.~S1 in Supplementary~\cite{Supplement}). 
The MI output is split into two paths ending with single-photon detectors. 
Measuring the coincidence count rate depending on the MI delay $\tau$ allows obtaining an interferogram:

\begin{figure}[h]
\centering\includegraphics[width=0.5\linewidth]{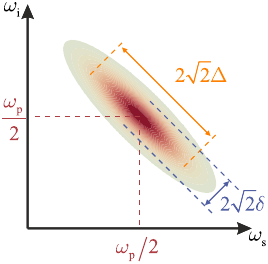}
\caption{Double-Gaussian two-photon spectral distributions $\abs{f(\omega_\text{s},\omega_\text{i})}$.}
\label{fig:Two_Omega}
\end{figure}

\begin{equation}
    \label{1.Mt} M(\tau)=\frac{1}{16}\left[ M_\text{c}+M_0(\tau)-M_1(\tau)+M_2(\tau)\right],
\end{equation}
\begin{subequations}
\begin{align}
\label{1.McM} \text{where}  \quad {{M}_{c}}&=\left(1+R\right)^2,\\
\label{1.M0}{{M}_{0}}(\tau)&=2R\ee^{ -\frac{{{\left( T-\tau  \right)}^{2}}}{2{{\Delta }^{-2}}}},\\
\label{1.M1}{{M}_{1}}(\tau)&=4r\left( 1+R \right)\ee^{ -\frac{{{\left( T-\tau  \right)}^{2}}}{8 \Delta _+^{-2}}}\cos \left[ {{\omega }_{0}}\left( T-\tau  \right) \right],\\
\label{1.M2}{{M}_{2}}(\tau)&=2R\ \ee^ { -\frac{{{\left( T-\tau  \right)}^{2}}}{2{{\delta }^{-2}}}}\cos \left[ 2{{\omega }_{0}}\left( T-\tau  \right) \right]. 
\end{align}
\end{subequations}
Here $\Delta_+^2\equiv\delta^2+\Delta^2\approx\Delta^2$ and $r^2\equiv R$ is a sample reflectivity.

\begin{figure}[ht]
\centering
\includegraphics[width=0.9\linewidth]{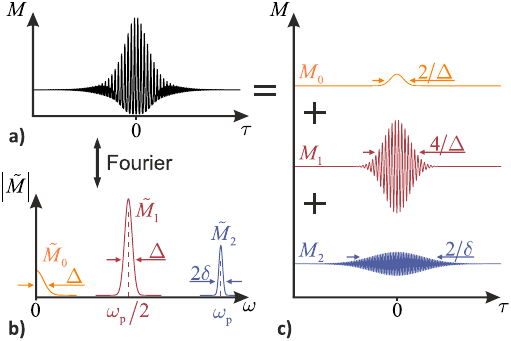}
\caption{Total interferogram $ M(\tau)$ (a), 
it's spectrum (b) with terms $\Tilde{M}_0 (\omega)$, $\Tilde{M}_1 (\omega)$, $\Tilde{M}_2 (\omega)$ and corresponding interferogram
terms $M_0 (\tau)$, $M_1 (\tau)$, $M_2 (\tau)$ (c).}
\label{fig:TheorPlots}
\end{figure}

The total interferogram $ M(\tau)$ is plotted in Fig.~\ref{fig:TheorPlots}a. 
Its spectrum $\abs{\Tilde{M}(\omega)}$ in Fig.~\ref{fig:TheorPlots}b contains three peaks: $\Tilde{M}_0$ centered at $\omega=0$ with StD $\Delta$, $\Tilde{M}_1$ centered at $\omega_0$ with StD $\Delta/2$, and $\Tilde{M}_2$ centered at $2 \omega_0$ with StD $\delta$. The corresponding interferogram terms are plotted in Fig.~\ref{fig:TheorPlots}c. 
Term $M_0$ represents a peak with StD $1/\Delta$, corresponding to HOM-interference. 
Term $M_1$ has twice the StD and corresponds to single-photon interference. 
Finally, $M_2$ term with a $1/\delta$ StD is equivalent to pump interference in an MI.   

The filtering of the $M_0$ term allows achieving a resolution that is twice as good as that of the $M_1$ term, corresponding to standard OCT. 
Moreover, similar to the usual HOM-based QOCT, the $M_0$ term provides dispersion cancellation (see Supplementary~\cite{Supplement} for details). 
However, we observe that terms $M_0$ and $M_1$ can be separated only if $\Delta\ll\omega_\text{p}/3$. 
Otherwise, the spectra $\Tilde{M}_0(\omega)$ and $\Tilde{M}_1(\omega)$ overlap, and the terms cannot be well separated anymore.

\subsection{Implementation}

For the basic experiment in QOCT, we constructed the setup illustrated in Fig.~\ref{fig:setup}b. Biphoton pairs were emitted from the fiber, collimated using a parabolic mirror.
The beam was then split on a non-polarized beam splitter (BS) into two channels: the sample channel and the reference channel. 
The sample channel included a mirror and an optional 2-mm wide glass plate to introduce dispersion. 
The reference channel consisted of a mirror mounted on a combined piezo-mechanical translator stage.
The reflected beams were recombined on the subsequent BS.

To broaden the detection bandwidth and cover both the visible (VIS) and infrared (IR) ranges, we employed a longpass dichroic mirror (DM) with a cut-on wavelength $\lambda_\text{c}=1000~\text{nm}$. 
In both channels, biphotons were
coupled with single-mode fibers using parabolic mirrors.
The infrared part was directed to an InGaAs avalanche
photodiode (APD) detector (MPD PDM-IR), labeled as
D1. The visible part was further split by a fiber BS and
directed to two Si-based APD detectors (Laser Components
COUNT NIR), labeled as D2 and D3. All detector
outputs were connected to coincidence circuits, enabling
coincidence measurements between the two visible-range
detectors D2\&D3 (VIS-VIS) and between the IR detector
and both visible-range detectors: D1\&D2 and D1\&D3
(IR-VIS).

\section{Experimental results and discussion}\label{sec:results}

\begin{figure*}[htbp]
\centering
\includegraphics[width=0.9\linewidth]{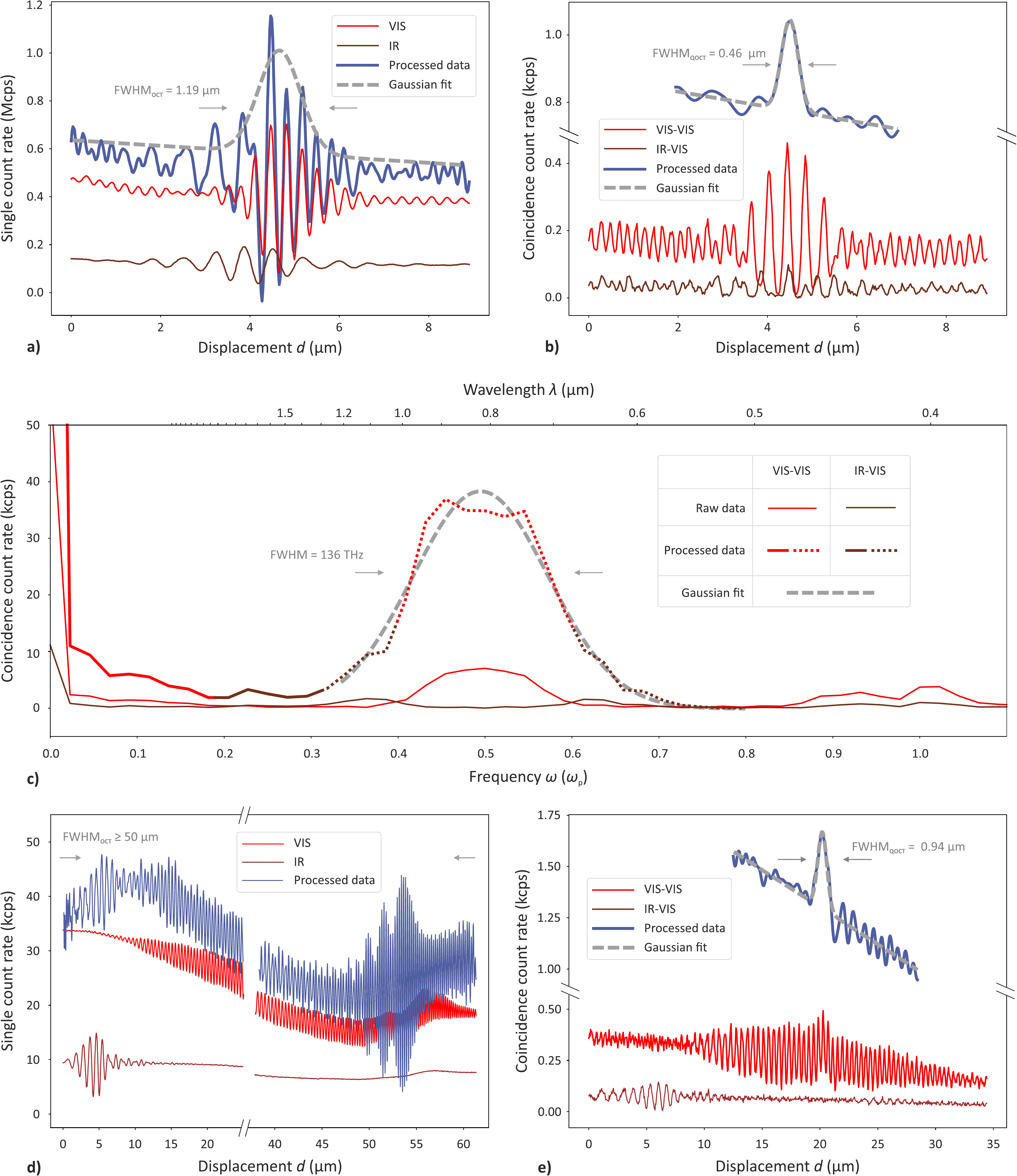}
\caption[font=\columnwidth, font=small]{Experimental results.
a) Interferograms in single count rate; the red curve refers to the raw data for wavelengths $400-1000$~nm (VIS), registered by Si APD detectors, the brown curve refers to the raw data for wavelengths $1000-1500$~nm (IR), registered by InGaAs APD detectors, the blue curve refers to the processed data where detection efficiency is taken into account and signals are summed, and the grey dotted curve corresponds to the Gaussian approximation of the processed data.
b) Interferograms in coincidence count rate; the red curve refers to the coincidences between two Si APDs (VIS-VIS), the brown curve shows coincidences between Si and InGaAs APDs (IR-VIS), the blue curve shows the processed data, and the grey dashed curve corresponds to the Gaussian fit. 
c) Coincidence spectra (absolute values of the Fast-Fourier Transform, FFT): the red thin line represents the raw VIS-VIS coincidence spectra, the brown thin line shows the spectrum of the raw IR-VIS coincidences, the bold red and brown solid lines correspond to the processed data used to extract the HOM term $M_0$, the dotted red and brown lines represent the processed spectrum used to extract the $\Tilde{M}_1$ term, and the grey dashed curve shows its Gaussian fit.
d) Interferograms in single count rate with dispersive media; the red curve refers to the raw VIS data, the brown curve refers to the raw IR data, and the blue curve refers to the processed data. 
e) Interferograms in coincidence count rate with dispersive media; the red curve refers to the raw VIS-VIS data, the brown curve refers to the raw IR-VIS data, the blue curve refers to the processed data, and the grey dashed line refers to the Gaussian fit.}
\label{fig:data}
\end{figure*}

Initially, we conducted a $z$-scan using a sample consisting only of a mirror. The results are depicted in Fig.~\ref{fig:data}a~--~c.
In Fig.~\ref{fig:data}a, we show raw single count rate interferograms for the VIS and IR channels (red and brown curves).
A small shift between their envelopes indicates slight dispersion in either the sample or reference arm.
Raw interferograms for VIS-VIS and IR-VIS coincidences are plotted in Fig.~\ref{fig:data}b, with the red and brown curves corresponding to each.

The absolute values of their Fast-Fourier Transform~(FFT) spectra are shown in Fig.~\ref{fig:data}c as thin curves with the same colors. The raw spectra qualitatively match the theory presented in Fig.~\ref{fig:TheorPlots}b (see the detailed analysis in Supplementary~\cite{Supplement}). They exhibit three peaks corresponding to the terms $\Tilde{M}_0$, $\Tilde{M}_1$, and $\Tilde{M}_2$, but due to the high bandwidth, the $\Tilde{M}_0$ and $\Tilde{M}_1$ peaks are partially overlapped. To extract the $\Tilde{M}_0$ term, we used the spectral range $0 - \omega_\text{p}/3$, and the range $\omega_\text{p}/3 - 3 \omega_\text{p}/4$ for extracting the $\Tilde{M}_1$ term.
To account for the spectral dependencies of the detectors' quantum efficiencies, $\eta_\text{VIS} (\omega)$ and $\eta_\text{IR} (\omega)$, we divided the raw spectra by the corresponding efficiency products (see details in Supplementary~\cite{Supplement}).

The processed spectrum for the $\Tilde{M}_1$ term, combined from both VIS-VIS and IR-VIS data, is plotted as a bold dotted curve. Its Gaussian fit, shown as a gray dashed curve, has a bandwidth $\text{FWHM}_\omega=2\pi\times 136~\text{THz}$.
Taking into account possible spectral broadening caused by interferometer instability (see details in Supplementary~\cite{Supplement}), the actual bandwidth value could be slightly lower, $\text{FWHM}_\omega^\text{(corrected)} = 2\pi \times 132~\text{THz}$, corresponding to a wavelength range of approximately 687--986~nm. This bandwidth is close to the estimated theoretical value of $\sqrt{\frac{4 \ln 2}{\pi}} B_\omega=2\pi\times117$~THz~\eqref{eq:2A_integral}, calculated for the measured pump waist $W=5.7\ \upmu \rm m$.

\afterpage{\clearpage}

Considering the spectral efficiency of the detectors and the bandwidth of coincidence detection (detailed information is provided in the Supplementary~\cite{Supplement}), the registered coincidence count rate of 3~kcps suggests a total source generation rate of $2.7 \   \frac{\text{kcps}}{\text{mW}}$. 
The source's spectral generation rate of $S_0 \approx 20~\frac{\text{cps}}{\text{THz}\times\text{mW}}$ is about 6 times smaller than the theoretical value of $125~\frac{\text{cps}}{\text{THz}\times\text{mW}}$~\cite{katamadze_broadband_2016}. 
The mismatch between the experimental and theoretical results can be explained by various aberrations and the wavelength dependency of the target mode waist, which were not accounted for in the theory. 
However, this efficiency was sufficient to achieve a registered coincidence count rate of about 25--30~$\frac{\text{cps}}{\text{mW}}$ in the raw interferogram, whereas previous works reported results of 2--4~$\frac{\text{cps}}{\text{mW}}$ for similar QOCT experiments~\cite{lopez-mago_quantum-optical_2012, okano_054_2015}.

Similarly, we divided the coincidence spectra related to the $\Tilde{M}_0$ term by the products of quantum efficiencies. The result is shown in Fig.~\ref{fig:data}c as thick solid curves. Its inverse FFT, corresponding to the HOM peak with an additional sine term $M_1$, is shown in Fig.~\ref{fig:data}b as a blue curve. The FWHM of its Gaussian fit (gray dashed curve), which defines the QOCT resolution, was estimated to be approximately $0.46~\upmu \rm m$, slightly better than the record QOCT resolution of $0.54~\upmu \rm m$ reported in~\cite{okano_054_2015}. This value is consistent with the value $2 \ln (2)c/\text{FWHM}_\omega=0.50 \ \upmu$m, corresponding to the measured spectral bandwidth.

We processed the single count rate interferograms in a similar manner, and the results are plotted in Fig.~\ref{fig:data}a as a blue curve. The FWHM of its Gaussian envelope (gray dashed curve), which defines the classical OCT resolution, was estimated to be $1.19~\upmu \rm m$. This is more than twice the QOCT resolution due to the presence of slight dispersion.

As a second step, we conducted a $z$-scan using a sample consisting of a mirror and a 2-mm glass plate, which introduces dispersion. The results are depicted in Fig.~\ref{fig:data}d--e.
The single-photon count rate interferograms (Fig.~\ref{fig:data}d) show well-separated IR and VIS terms due to dispersion. The dispersion-induced broadening was significant enough that a single scan couldn't cover the interference pattern range, requiring two scans to estimate the bandwidth ($\approx 50~\upmu\rm m$).

The raw coincidence interferograms and the filtered HOM term, accounting for detector efficiencies, are shown in Fig.~\ref{fig:data}e. The obtained resolution under such significant dispersion was $0.94~\upmu \rm m$, which is 50 times better than the OCT resolution but still twice as large as the QOCT resolution without dispersion. The non-ideal dispersion cancellation can be attributed to the influence of higher-order dispersion.

\section{Conclusion}\label{sec:conclusion}
We have introduced a cost-effective technique for generating a bright (2.7~kcps/mW pair generation rate) and broadband (132~THz bandwidth) biphoton field using a bulk nonlinear crystal with tight pump beam focusing. 
Demonstrating both theoretical insights and experimental validation, we applied this technique to quantum optical coherence tomography (QOCT), achieving an impressive axial resolution of $0.46 \ \upmu$m, comparable to the record results reported in~\cite{okano54MmResolution2015}. 
Significantly, we have demonstrated for the first time that Michelson interferometer-based QOCT exhibits dispersion cancellation, analogous to that observed in traditional Hong-Ou-Mandel-based implementations, resulting in at least a 50-fold improvement in resolution compared to classical Optical Coherence Tomography technique. 
Furthermore, our results suggest that the achieved QOCT resolution is limited by interferogram filtering procedures, indicating potential for further enhancement through the exploration of advanced filtering methods or modified interference schemes.

\begin{acknowledgments}
The authors would like to thank the anonymous referees for their insightful questions and constructive comments, which helped to significantly improve the quality of this work.
\end{acknowledgments}

\bibliography{apssamp}

\end{document}


\title{Broadband biphoton source for quantum optical coherence tomography \\ based on a Michelson interferometer: supplemental material}


\maketitle

\tableofcontents

\section{Biphoton interference in a Michelson interferometer}
\subsection{Main description}
\begin{figure}
    \centering
    \includegraphics[width=0.5 \textwidth]{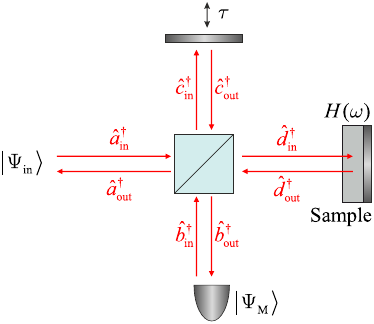}
    \caption{Michelson Interferometer.}
    \label{fig:Michelson}
\end{figure}
We describe the interferometer following the works
\cite{odateTwophotonQuantumInterference2005, lopez-mago_coherence_2012, lopez-mago_quantum-optical_2012}.
Consider a Michelson interferometer, as illustrated in Fig.~\ref{fig:Michelson}. It comprises a 50/50 beam splitter, a reflective sample, and a mirror in a reference channel. It connects four input and four output optical modes. Photon creation operators in these modes are denoted as $\hat{a}^\dagger_{\text{in}}$, $\hat{a}^\dagger_{\text{out}}$, $\hat{b}^\dagger_{\text{in}}$, $\hat{b}^\dagger_{\text{out}}$, $\hat{c}^\dagger_{\text{in}}$, $\hat{c}^\dagger_{\text{out}}$, $\hat{d}^\dagger_{\text{in}}$, $\hat{d}^\dagger_{\text{out}}$. The beam splitter connects these operators as follows:
\begin{equation}\label{Michelson1.matrices1}
   \left( \begin{matrix}
   \hat{a}_{\text{in}}^{\dagger }  \\
   \hat{b}_{\text{in}}^{\dagger }  \\
\end{matrix} \right)=\frac{1}{\sqrt{2}}\left( \begin{matrix}
   1 & \text{i}  \\
   \text{i} & 1  \\
\end{matrix} \right)\left( \begin{matrix}
   \hat{d}_{\text{in}}^{\dagger }  \\
   \hat{c}_{\text{in}}^{\dagger }  \\
\end{matrix} \right)
,\quad
\left( \begin{matrix}
   \hat{c}_{\text{out}}^{\dagger }  \\
   \hat{d}_{\text{out}}^{\dagger }  \\
\end{matrix} \right)=\frac{1}{\sqrt{2}}\left( \begin{matrix}
   1 & -\text{i}  \\
   -\text{i} & 1  \\
\end{matrix} \right)\left( \begin{matrix}
   \hat{a}_{\text{out}}^{\dagger }  \\
   \hat{b}_{\text{out}}^{\dagger }  \\
\end{matrix} \right).
\end{equation}
Denote the sample response function as $H(\omega)$ and delay in the reference channel as $\tau$. Then input and output modes are connected as follows:
\begin{equation}\label{Michelson1.matrices2}
    \left( \begin{matrix}
   \hat{c}_{\text{in}}^{\dagger }  \\
   \hat{d}_{\text{in}}^{\dagger }  \\
\end{matrix} \right)=\left( \begin{matrix}
   {{\operatorname{e}}^{\operatorname{i}\omega \tau }} & \text{0}  \\
   \text{0} & H\left( \omega  \right)  \\
\end{matrix} \right)\left( \begin{matrix}
   \hat{c}_{\text{out}}^{\dagger }  \\
   \hat{d}_{\text{out}}^{\dagger }  \\
\end{matrix} \right).
\end{equation}
Combining equations \eqref{Michelson1.matrices1} and \eqref{Michelson1.matrices2}, we obtain:
\begin{equation}\label{Michelson1.operators}
\hat{a}_{\text{in}}^{\dagger }=\frac{1}{2}\left( H\left( \omega  \right)+{{\operatorname{e}}^{\operatorname{i}\omega \tau }} \right)\hat{a}_{\text{out}}^{\dagger }+\frac{\operatorname{i}}{2}\left( H\left( \omega  \right)-{{\operatorname{e}}^{\operatorname{i}\omega \tau }} \right)\hat{b}_{\text{out}}^{\dagger }.
\end{equation}

Consider an input collinear biphoton state:
\begin{equation}\label{Michelson1.Psiin}
    \left| {{\Psi }_{\text{in}}} \right\rangle =
    \frac{1}{\sqrt{2}}
    \iint{\text{d}{{\omega }_{s}}\text{d}{{\omega }_{i}}}f\left( {{\omega }_{s}},{{\omega }_{i}} \right)\hat{a}_{\text{in}}^{\dagger }\left( {{\omega }_{s}} \right)\hat{a}_{\text{in}}^{\dagger }\left( {{\omega }_{i}} \right)\left| \text{vac} \right\rangle, 
\end{equation}
where $f\left( {{\omega }_{s}},{{\omega }_{i}} \right)$ is a spectral amplitude. Assume it is symmetrical, so that $f\left( {{\omega }_{s}},{{\omega }_{i}} \right)=f\left( {{\omega }_{i}},{{\omega }_{s}} \right)$, which is true for type-I SPDC.

Substituting \eqref{Michelson1.operators} into \eqref{Michelson1.Psiin} we obtain
\begin{equation}\label{Michelson1.Psiin1}
\left| {{\Psi }_{\text{out}}} \right\rangle=
-\frac{1}{{4\sqrt{2}}}
\iint{\text{d}{{\omega }_{s}}\text{d}{{\omega }_{i}}}
f_{\text{out}}\left( {{\omega }_{s}},{{\omega }_{i}} \right)
    \hat b^\dagger_{\text{out}}(\omega_{\text{s}})
    \hat b^\dagger_{\text{out}}(\omega_{\text{i}})
    \left| \text{vac} \right\rangle+\cdots
\end{equation}
where
\begin{equation}\label{Michelson1.Psiin2}
f_{\text{out}}\left( {{\omega }_{s}},{{\omega }_{i}} \right)\equiv
f\left( {{\omega }_{s}},{{\omega }_{i}} \right)
    \left[
    H(\omega_{\text{s}})-\ee^{\ii\omega_{\text{s}}\tau}
    \right]
    \left[
    H(\omega_{\text{i}})-\ee^{\ii\omega_{\text{i}}\tau}
    \right]
\end{equation}

Here we are interested only in the term where we have two photons in the output mode $b$ because we consider the detector realising projection on the two-photon state:
\begin{equation}\label{Michelson1.PsiM}
    \left| {{\Psi }_{M}}\left( {{\omega }_{1}},{{\omega }_{2}} \right) \right\rangle =\frac{1}{\sqrt{2}}\hat{b}_{\text{out}}^{\dagger }\left( {{\omega }_{1}} \right)\hat{b}_{\text{out}}^{\dagger }\left( {{\omega }_{2}} \right)\left| \text{vac} \right\rangle.
\end{equation}
Therefore the probability to register two photons, depending on the delay in the reference channel (two-photon interferogram) is given by:
\begin{equation}\label{Michelson1.iterferogram1}
    M\left( \tau  \right)=\iint{\text{d}{{\omega }_{1}}\text{d}{{\omega }_{2}}}S\left( {{\omega }_{1}},{{\omega }_{2}} \right),\quad \text{where} \quad S\left( {{\omega }_{1}},{{\omega }_{2}} \right)\equiv {{\left| \left\langle  {{\Psi }_{M}}\left( {{\omega }_{1}},{{\omega }_{2}} \right) | {{\Psi }_{\text{out}}} \right\rangle  \right|}^{2}}.
\end{equation}
The projection is given by 
\begin{align}\nonumber
\left\langle  {{\Psi }_{M}}\left( {{\omega }_{1}},{{\omega }_{2}} \right) | {{\Psi }_{\text{out}}} \right\rangle 
&=
\left[ {{f}_{\text{out}}}\left( {{\omega }_{1}},{{\omega }_{2}} \right)+{{f}_{\text{out}}}\left( {{\omega }_{2}},{{\omega }_{1}} \right) \right]\\
&=
\frac{1}{4} f\left( {{\omega }_{1}},{{\omega }_{2}} \right)
    \left[
    H(\omega_1)-\ee^{\ii\omega_1\tau}
    \right]
    \left[
    H(\omega_2)-\ee^{\ii\omega_2\tau}
    \right].
\end{align}

Let's make a change of variables: $\omega_1=\omega_0+\nu_1$, $\omega_2=\omega_0+\nu_2$, where $\omega_0\equiv\omega_{\text{p}}/2$ -- half of the pump frequency.
In this case, we have
\begin{equation}
    \abs{
 \left[
    H(\omega_0+\nu_1)-\ee^{\ii(\omega_0+\nu_1)\tau}
    \right]
    \left[
    H(\omega_0+\nu_2)-\ee^{\ii(\omega_0+\nu_2)\tau}
    \right]
 }^2=K_c+K_0-K_1+K_2,
\end{equation}
%
where
\begin{subequations}
\begin{align}
K_c\equiv &\left(\abs{H(\omega_0+\nu_1)}^2+1\right)\left(\abs{H(\omega_0+\nu_2)}^2+1\right)\label{Michelson1.Kc} ,\\
K_0\equiv & 2\Re\left[ \ee^{-\ii(\nu_1-\nu_2)\tau} H(\omega_0+\nu_1)H^\ast(\omega_0+\nu_2) \right]\label{Michelson1.K0},\\ \nonumber
K_1\equiv & 2\Re\left[\ee^{-\ii\omega_0\tau} \ee^{-\ii\nu_1\tau}H(\omega_0+\nu_1)\left(\abs{H(\omega_0+\nu_2)}^2+1\right)\right]+\\
&2\Re\left[\ee^{-\ii\omega_0\tau} \ee^{-\ii\nu_2\tau}H(\omega_0+\nu_2)\left(\abs{H(\omega_0+\nu_1)}^2+1\right)\right]\label{Michelson1.K1},\\
K_2\equiv& 2\Re\left[\ee^{-2\ii\omega_0\tau}\ee^{-\ii(\nu_1+\nu_2)\tau}H(\omega_0+\nu_1)H(\omega_0+\nu_2)
\right].\label{Michelson1.K2}
\end{align}
\end{subequations}
Therefore the final interferogram consists of four terms:
\begin{equation}\label{Michelson1.FinalInterferogram}
   16 M(\tau)=M_c^{(M)}+M_0(\tau)-M_1(\tau)+M_2(\tau),
\end{equation}
where
\begin{subequations}\label{0M}
\begin{align}
    {{M}_{c}}&=\iint{\text{d}{{\nu }_{1}}\text{d}{{\nu }_{2}}}{{\left| f\left( {{\omega }_{0}}+{{\nu }_{1}},{{\omega }_{0}}+{{\nu }_{2}} \right) \right|}^{2}}{{K}_{c}^{(M)}}\quad -\quad \text{Constant},\label{Michelson0.Mc}\\
    {{M}_{0}}(\tau)&=\iint{\text{d}{{\nu }_{1}}\text{d}{{\nu }_{2}}}{{\left| f\left( {{\omega }_{0}}+{{\nu }_{1}},{{\omega }_{0}}+{{\nu }_{2}} \right) \right|}^{2}}{{K}_{0}}\quad -\quad \text{HOM peak},\label{Michelson0.M0}\\
    {{M}_{1}}(\tau)&=\iint{\text{d}{{\nu }_{1}}\text{d}{{\nu }_{2}}}{{\left| f\left( {{\omega }_{0}}+{{\nu }_{1}},{{\omega }_{0}}+{{\nu }_{2}} \right) \right|}^{2}}{{K}_{1}}\quad -\quad \text{Single-photon interference},\label{Michelson0.M1}\\
    {{M}_{2}}(\tau)&=\iint{\text{d}{{\nu }_{1}}\text{d}{{\nu }_{2}}}{{\left| f\left( {{\omega }_{0}}+{{\nu }_{1}},{{\omega }_{0}}+{{\nu }_{2}} \right) \right|}^{2}}{{K}_{2}}\quad -\quad \text{Pump interference}.\label{Michelson0.M2}
\end{align}
\end{subequations}

\subsection{Examples}
In this section, we explore two types of two-photon spectral distributions $\abs{f(\omega_{\text{s}},\omega_{\text{i}})}$: degenerate and non-degenerate. Additionally, we examine two types of sample response functions $H(\omega)$: one without dispersion and another with dispersion.

\subsubsection{Degenerate SPDC, without Dispersion}
\begin{figure}
    \centering
    \includegraphics[width=0.5 \textwidth]{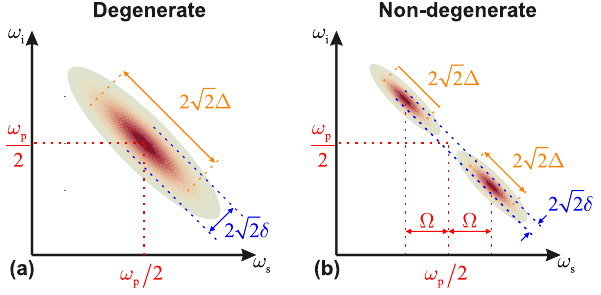}
    \caption{Two-photon spectral distributions $\abs{f(\omega_{\text{s}},\omega_{\text{i}})}$ for degenerate (a) and non-degenerate (b) cases.}
    \label{fig:TwoOmega}
\end{figure}

We first consider a normalized double-Gaussian two-photon spectrum for degenerate SPDC, visualized in Fig.~\ref{fig:TwoOmega}a~\cite{Mikhailova2008,Fedorov2009}:


\begin{equation}
\label{degenerate} 
  {{\left| f({{\omega }_{s}},{{\omega }_{i}}) \right|}^{2}}=\frac{1}{\pi \delta \Delta }\underbrace{\exp \left[ -\frac{{{\left( {{\omega }_{1}}-{{\omega }_{2}} \right)}^{2}}}{2{{\Delta }^{2}}} \right]}_{\text{Phasematching}}\underbrace{\exp \left[ -\frac{{{\left( {{\omega }_{1}}+{{\omega }_{2}}-2{{\omega }_{0}} \right)}^{2}}}{2{{\delta }^{2}}} \right]}_{\text{Pump}},
\end{equation}
where $\int
{{{\left| f({{\omega }_{s}},{{\omega }_{i}}) \right|}^{2}}\text{d}{{\omega }_{s}}\text{d}{{\omega }_{i}}}=1$.
Here, $2\delta$ represents the pump bandwidth, and $2 \Delta$ is the phasematching bandwidth (here and bellow the width of spectral and temporal distributions is defined as a double standard deviation). 

Consider a single-layer sample without dispersion with the following response function:
\begin{equation}\label{single}
    H\left( \omega  \right)=r{{\operatorname{e}}^{\operatorname{i}\omega T}},\quad {{r}^{2}}\equiv R - \text{layer reflectivity}.
\end{equation}
Here $\omega T = k d=\frac{n \omega d}{c}$, where $d$ is the sample thickness, $n$ is the refractive index and $c$ is the speed of light. So $T=\frac{nd}{c}$. 

\begin{figure}[b]
    \centering
    \includegraphics[width= \textwidth]{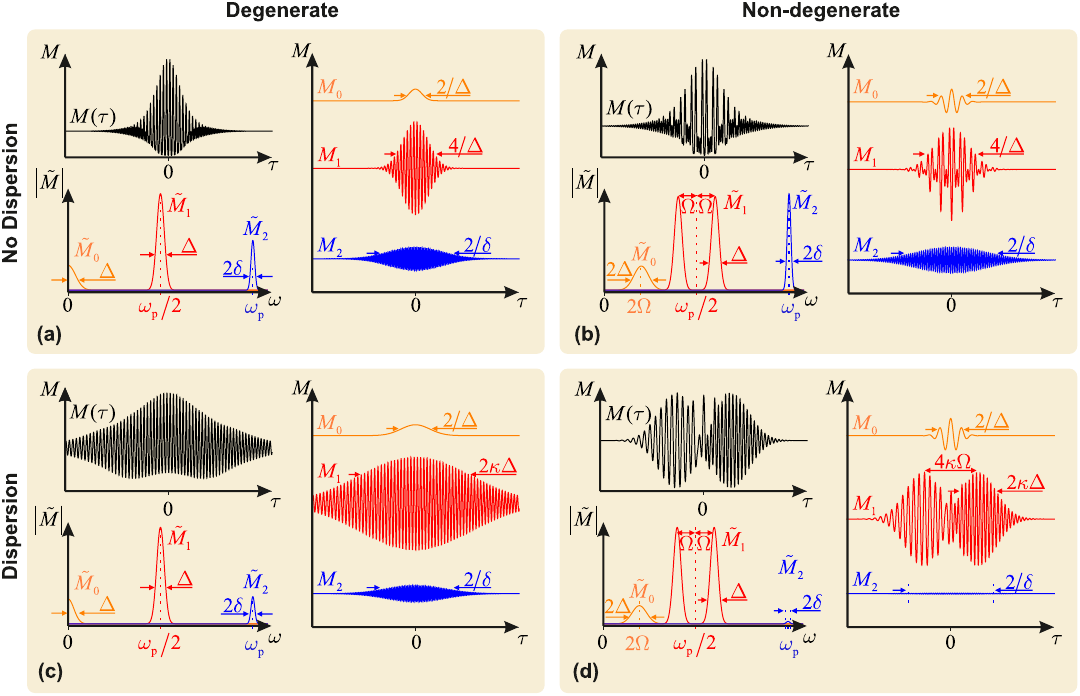}
    \caption{Full interferograms $M(\tau)$, their spectra $\abs{\Tilde{M}(\omega)}$, different spectral components $\Tilde{M}_0 (\omega)$, $\Tilde{M}_1 (\omega)$, $\Tilde{M}_2 (\omega)$ and corresponding interferogram terms $M_0(\tau)$, $M_1(\tau)$, $M_2(\tau)$ for degenerate and non-degenerate two-photon spectral distributions $\abs{f(\omega_{\text{s}},\omega_{\text{i}})}$ and for two cases of sample response functions $H(\omega)$: with and without optical dispersion.  }
    \label{fig:TheorPlots}
\end{figure}

Substituting (\ref{degenerate},\ref{single}) into \eqref{0M}, 
we obtain:
\begin{subequations}\label{1M}
\begin{align}
\label{1.McM}{{M}_{c}}&=\left(1+R\right)^2,\\
\label{1.M0}{{M}_{0}}(\tau)&=2R\exp \left[ -\frac{{{\left( T-\tau  \right)}^{2}}}{2{{\Delta }^{-2}}} \right],\\
\label{1.M1}{{M}_{1}}(\tau)&=4r\left( 1+R \right)\exp \left[ -\frac{\Delta _{+}^{2}{{\left( T-\tau  \right)}^{2}}}{8} \right]\cos \left[ {{\omega }_{0}}\left( T-\tau  \right) \right],\\
\label{1.M2}{{M}_{2}}(\tau)&=2R\ \exp \left[ -\frac{{{\left( T-\tau  \right)}^{2}}}{2{{\delta }^{-2}}} \right]\cos \left[ 2{{\omega }_{0}}\left( T-\tau  \right) \right], 
\end{align}
\end{subequations}
where $\Delta_+^2=\delta^2+\Delta^2\approx\Delta^2$.
The total interferogram $ M(\tau)=\frac{1}{16}\left[ M_c+M_0(\tau)-M_1(\tau)+M_2(\tau)\right]$, its terms $M_0 (\tau)$, $M_1 (\tau)$, $M_2 (\tau)$ and their spectra $\Tilde{M}_0 (\omega)$, $\Tilde{M}_1 (\omega)$, $\Tilde{M}_2 (\omega)$ are plotted in Fig.~\ref{fig:TheorPlots}a. 
Notice that all the terms are spectrally separated: $\Tilde{M}_0 (\omega)$ centered at $\omega=0$ with a bandwidth $2 \Delta$, $\Tilde{M}_1 (\omega)$ centered at $\omega=\omega_0=\omega_{\text{p}}/2$ with a bandwidth $\Delta$, $\Tilde{M}_2 (\omega)$ centered at $\omega=2\omega_0=\omega_{\text{p}}$ with a bandwidth $2\delta$. 
Thus, each term can be filtered. Term $M_0(\tau)$ corresponds to HOM interference and has a temporal width $2/\Delta$, $M_1(\tau)$ corresponds to single-photon Michelson interference and has a width $4/\Delta$ and $M_2(\tau)$ corresponds to double-photon (pump) Michelson interference and has a width $2/\delta$. 
Therefore, filtering the $M_0$ term one can achieve twice the resolution compared to $M_1$, which corresponds to standard OCT. 
However, it is observed  that the terms $M_0$ and $M_1$ can be separated effectively only if $\Delta\ll\omega_{\text{p}}/3$. 
Otherwise, the spectra $\Tilde{M}_0(\omega)$ and $\Tilde{M}_1(\omega)$ overlap, and the terms cannot be well-separated anymore. 

\subsubsection{Non-degenerate SPDC, without Dispersion}

Consider now the case of non-degenerate SPDC with the following two-photon spectral distribution, illustrated in Fig.~\ref{fig:TwoOmega}b:
\begin{equation}\label{nondegenerate}
\begin{gathered}
    {{\left| f({{\omega }_{s}},{{\omega }_{i}}) \right|}^{2}}=\frac{1}{2\pi \delta \Delta }\underbrace{\left( \exp \left[ -\frac{{{\left( {{\omega }_{1}}-{{\omega }_{2}}-2\Omega  \right)}^{2}}}{2{{\Delta }^{2}}} \right]+\exp \left[ -\frac{{{\left( {{\omega }_{1}}-{{\omega }_{2}}+2\Omega  \right)}^{2}}}{2{{\Delta }^{2}}} \right] \right)}_{\text{Phasematching}}\times\\
    \underbrace{\exp \left[ -\frac{{{\left( {{\omega }_{1}}+{{\omega }_{2}}-2{{\omega }_{0}} \right)}^{2}}}{2{{\delta }^{2}}} \right]}_{\text{Pump}}.
\end{gathered}
\end{equation}
Here, $\Omega$ is a frequency detuning of the central frequencies. Such a distribution can be obtained directly through phase matching tuning or due to postselection by placing some filters before the detectors. 
In our experiment, measuring coincidences between Si and InGaAs APDs resulted in spectral response functions shifted from the central frequency $\omega_0$, equivalent to postselecting a biphoton field with a non-degenerate spectrum.


Consider again the single-layer refractive sample without dispersion \eqref{single}.
Substituting (\ref{degenerate},\ref{single}) into \eqref{0M}, we obtain:
\begin{subequations}\label{2M}
\begin{align}
\label{2.McM}{{M}_{c}}&=\left(1+R\right)^2\\
\label{2.M0}{{M}_{0}}(\tau)&=2R\,\exp \left[ -\frac{{{\left( T-\tau  \right)}^{2}}}{2{{\Delta }^{-2}}} \right]\cos \left[ 2 \Omega \left( T-\tau  \right) \right] \\
\label{2.M1}{{M}_{1}}(\tau)&=4r\left( 1+R \right)
\exp 
\left[ -\frac{\Delta _{+}^{2}{{\left( T-\tau  \right)}^{2}}}{8} \right]
\cos \left[ {{\omega }_{0}}\left( T-\tau  \right) \right]
\cos \left[ \Omega \left( T-\tau  \right) \right] \\
\label{2.M2}{{M}_{2}}(\tau)&=2R\ \exp \left[ -\frac{{{\left( T-\tau  \right)}^{2}}}{2{{\delta }^{-2}}} \right]\cos \left[ 2{{\omega }_{0}}\left( T-\tau  \right) \right].
\end{align}
\end{subequations}

The total interferogram $M(\tau)$, it's terms $M_0 (\tau)$, $M_1 (\tau)$, $M_2 (\tau)$ and their spectra $\Tilde{M}_0 (\omega)$, $\Tilde{M}_1 (\omega)$, $\Tilde{M}_2 (\omega)$ are plotted in Fig.~\ref{fig:TheorPlots}b. All the terms again can be spectrally separated: $\Tilde{M}_0 $ has a broadband peak ($2\Delta$ bandwidth) centered at $2\Omega$, $\Tilde{M}_1$ has two peaks ($\Delta$ bandwidth) centered at $\omega_{\text{p}}/2\pm\Omega$, and $\Tilde{M}_2$ has a narrowband peak ($2\delta$ bandwidth) centered at $\omega_{\text{p}}$. Again, $M_0 (\tau)$ interferogram term is two times narrower than the standard OCT term $M_1(\tau)$, and it can be spectrally filtered if $2\Omega+\Delta\ll \omega_{\text{p}}/3$.

\subsubsection{Degenerate SPDC, with Dispersion}

Consider the sample with dispersion, where the phase shift $kd=\frac{n\left( \omega  \right)\omega d}{c}$. The refractive index dispersion can be described as $n={{n}_{0}}+\eta \left( \omega -{{\omega }_{0}} \right)$, where $n_0\equiv n(\omega_0)$ and $\eta \equiv{{\left. \frac{\partial n}{\partial \omega } \right|}_{{{\omega }_{0}}}}$. 
So the phase shift can be expressed as
\begin{equation}
    kd=\frac{d}{c}\left[ {{n}_{0}}+\eta \left( \omega -{{\omega }_{0}} \right) \right]\omega =\underbrace{\frac{d}{c}\left( {{n}_{0}}+\eta {{\omega }_{0}} \right)}_{T}\omega +\underbrace{\frac{d}{c}\eta }_{\kappa }{{\left( \omega -{{\omega }_{0}} \right)}^{2}}-\underbrace{\frac{d}{c}\eta \omega _{0}^{2}}_{\text{const}}.
\end{equation}
Omiting the last constant term, we have the following sample response function:
\begin{equation}\label{dispersion}
    H\left( \omega  \right)=r{{\operatorname{e}}^{\operatorname{i}\left[ \omega T+\kappa {{\left( \omega -{{\omega }_{0}} \right)}^{2}} \right]}},\quad {{r}^{2}}=R,\quad \kappa =\frac{d}{c}{{\left. \frac{\partial n}{\partial \omega } \right|}_{{{\omega }_{0}}}}.
\end{equation}

Assume first that the input biphoton field has a degenerate spectral distribution \eqref{degenerate}. Then, by  substituting (\ref{degenerate},\ref{dispersion}) into \eqref{0M}, we obtain:
\begin{subequations}\label{3M}
\begin{align}
\label{3.McM}{{M}_{c}}=&\left(1+R\right)^2,\\
\label{3.M0}
M_0 ( \tau ) =&\frac{2 R }{\sqrt{1+\delta ^2 \Delta ^2 \kappa ^2}}
\exp \left(-\frac{\Delta ^2 (T-\tau )^2}{2+2 \delta ^2 \Delta ^2 \kappa ^2}\right) ,
\\  \nonumber
M_1 (\tau) =&-\frac{4 \sqrt{2} r (R+1) }{\sqrt[4]{\Delta _+^4 \kappa ^2+4}}
\exp \left(-\frac{\Delta _+^2 (T-\tau)^2}{2 \left(\Delta _+^4 \kappa ^2+4\right)}\right)
\times\\ \label{3.M1}
&
\cos \left(
\omega_0 (T-\tau )
-\frac{\Delta _+^4 \kappa  (T-\tau )^2}{4 \left(\Delta _+^4 \kappa ^2+4\right)}
-\frac{1}{2} \arg  \left(2-\ii \Delta _+^2 \kappa \right)
\right),
\\ \nonumber
M_2(\tau)=&\frac{R}{2} 
\exp \left(-\frac{\delta ^2 (T-\tau )^2}{2 \delta ^4 \kappa ^2+2}\right)
\Re\Bigg\{
\frac{\exp \left(
2 \ii \omega _0 (T-\tau )
-\frac{\ii \delta ^4 \kappa  (T-\tau )^2}{2 \left(\delta ^4 \kappa ^2+1\right)}
\right)}{\sqrt{\left(\delta ^2 (-\kappa )-\ii\right) \left(\Delta ^2 \kappa +\ii\right)}}\times\\ \label{3.M2}
&\left[
4+2 \text{erf}\left(\frac{\sqrt{\frac{\left(1-\ii \delta ^2 \kappa \right) \left(\Delta ^2 \kappa +\ii\right)}{\delta ^2 \left(4 \Delta ^2 \kappa +2 \ii\right)+2 \ii \Delta ^2}} \delta ^2 \left(\Delta ^2 \kappa +\ii\right) (T-\tau ) }{\left(\delta ^2 \kappa +\ii\right) \left(\Delta ^2 \kappa +\ii\right)}\right)
\right]
\Bigg\},
\end{align}
\end{subequations}
where $\text{erf}$ is the error function. Looking at the term $M_0$, by comparing \eqref{3.M0} and \eqref{1.M0}, one can note that dispersion leads to an increase in the  width of $M_0(\tau)$ by a factor of $1+\delta^2\Delta^2\kappa^2$.

Therefore, similar to conventional HOM-interferometer-based QOCT~\cite{okanoDispersionCancellationHighresolution2013}
the dispersion cancellation condition is $\delta^2\Delta^2\kappa^2\ll 1$.
Assuming this condition, a narrowband pump $\delta \ll \Delta$, and significant dispersion for standard OCT $\kappa\Delta^2\gg 1$, we can simplify \eqref{3M}: 
\begin{subequations}\label{4M}
\begin{align}
\label{4.McM}{{M}_{c}}=&\left(1+R\right)^2,\\
\label{4.M0}
M_0 ( \tau ) =&
2 R \exp \left(-\frac{\Delta ^2 (T-\tau )^2}{2 \Delta^{-2}}\right),\\
\label{4.M1}
M_1 (\tau) =&-\frac{8 r (R+1) }{\Delta \sqrt{2\kappa}}
\exp \left(-\frac{(T-\tau)^2}{2 \Delta^2 \kappa ^2}\right)
\cos \left[
\omega_0 (T-\tau )
-\frac{(T-\tau )^2}{4\kappa}
+\frac{\pi}{4}
\right],
\\
M_2(\tau)=&\frac{2 R}{\Delta\sqrt{\kappa}} 
\exp \left(-\frac{(T-\tau )^2}{2 \delta ^{-2}}\right)
\cos\left[
2\omega_0 (T-\tau)-\frac{\delta^4\kappa (T-\tau)^2}{2}+\frac{\pi}{4}
\right].
\end{align}
\end{subequations}

The total interferogram, its terms, and their spectra are plotted in Fig.~\ref{fig:TheorPlots}c. The spectra $\abs{\Tilde{M}(\omega)}$ look similar to the case without dispersion (Fig.~\ref{fig:TheorPlots}a). 
All the peak centers and bandwidths are the same, 
but the $\Tilde{M}_1$ peak is $\Delta\sqrt{\kappa}$  times lower.
Corresponding interferogram terms $M_0$ and $M_2$ looks similar to Fig.~\ref{fig:TheorPlots}a; just the amplitude of the $M_2$ term is $\Delta\sqrt{\kappa}$ times smaller. However, the $M_1$ term related to standard OCT is not Fourier-limited anymore, and its width is limited by dispersion and equals $2\kappa\Delta$. 
So, if $\Delta \ll \omega_{\text{p}}/3$, one can separate the $M_0$ term and obtain dispersion cancellation for MI-based QOCT.

\subsubsection{Non-degenerate SPDC, with Dispersion}

Finally, consider the combination of a non-degenerate input biphoton field \eqref{nondegenerate} and a dispersive sample \eqref{dispersion}. By substituting (\ref{nondegenerate},\ref{dispersion}) into \eqref{0M}, we obtain:
\begin{subequations}\label{5M}
\begin{align}
\label{5.McM}{{M}_{c}}=&\left(1+R\right)^2,\\
\label{5.M0}
M_0 ( \tau ) =&\frac{2 R }{\sqrt{1+\delta ^2 \Delta ^2 \kappa ^2}}
\exp \left(-\frac{4 \delta ^2 \kappa ^2 \Omega ^2+\Delta ^2 (T-\tau )^2}{2+2 \delta ^2 \Delta ^2 \kappa ^2}\right) 
\cos \left(\frac{2 \Omega  (T-\tau )}{1+\delta ^2 \Delta ^2 \kappa ^2}\right)
\\ \nonumber
M_1 (\tau) =&-\frac{2 \sqrt{2} r (R+1) }{\sqrt[4]{\Delta _+^4 \kappa ^2+4}}
\Bigg\{
\exp \left(-\frac{\Delta _+^2 (T-\tau + 2 \kappa  \Omega  )^2}{2 \left(\Delta _+^4 \kappa ^2+4\right)}\right) \times\\ \nonumber
&\cos \left(
\omega_0 (T-\tau )
+\frac{4 \Omega  (T-\tau )}{\Delta _+^4 \kappa ^2+4}
-\frac{\Delta _+^4 \kappa  (T-\tau )^2}{4 \left(\Delta _+^4 \kappa ^2+4\right)}
+\frac{4 \kappa  \Omega ^2}{\Delta _+^4 \kappa ^2+4}
-\frac{1}{2} \arg  \left(2-\ii \Delta _+^2 \kappa \right)
\right)+\\ \nonumber
&\exp \left(-\frac{\Delta _+^2 (T-\tau - 2 \kappa  \Omega)^2}{2 \left(\Delta _+^4 \kappa ^2+4\right)}\right)\times\\ \label{5.M1}
&\cos \left(
\omega_0 (T-\tau )
-\frac{4 \Omega  (T-\tau )}{\Delta _+^4 \kappa ^2+4}
+\frac{4 \kappa  \Omega ^2}{\Delta _+^4 \kappa ^2+4}
-\frac{\Delta _+^4 \kappa  (T-\tau )^2}{4 \left(\Delta _+^4 \kappa ^2+4\right)}
-\frac{1}{2} \arg  \left(2-\ii \Delta _+^2 \kappa \right)
\right)
\Bigg\},
\\ \nonumber
M_2(\tau)=&\frac{R}{2} 
\exp \left(-\frac{\Delta ^2 \kappa ^2 \Omega ^2}{2 \Delta ^4 \kappa ^2+2}-\frac{\delta ^2 (T-\tau )^2}{2 \delta ^4 \kappa ^2+2}\right)\times\\ \nonumber
&\Re\Bigg\{
\frac{\exp \left(
2 \ii \omega _0 (T-\tau )
-\frac{\ii \delta ^4 \kappa  (T-\tau )^2}{2 \left(\delta ^4 \kappa ^2+1\right)}
+\frac{2 \ii \kappa  \Omega ^2}{\Delta ^4 \kappa ^2+1}
\right)}{\sqrt{\left(\delta ^2 (-\kappa )-\ii\right) \left(\Delta ^2 \kappa +\ii\right)}}\times\\ \label{5.M2}
&\left[
4+2 \text{erf}\left(\frac{\sqrt{\frac{\left(1-\ii \delta ^2 \kappa \right) \left(\Delta ^2 \kappa +\ii\right)}{\delta ^2 \left(4 \Delta ^2 \kappa +2 \ii\right)+2 \ii \Delta ^2}} \left(2 \Omega +\delta ^2 \left(\left(\Delta ^2 \kappa +\ii\right) (T-\tau )-2 \ii \kappa  \Omega \right)\right)}{\left(\delta ^2 \kappa +\ii\right) \left(\Delta ^2 \kappa +\ii\right)}\right)
\right]
\Bigg\}.
\end{align}
\end{subequations}
Again,  considering the assumptions 
of dispersion cancellation ${{\delta }^{2}}{{\Delta }^{2}}{{\kappa }^{2}}\ll 1$, a narrowband pump  ${{\delta }}\ll {\Delta }\sim\Omega$, and dispersion significance $\kappa {{\Delta }^{2}}\gg 1$,
we simplify \eqref{5M}: 
\begin{subequations}
\begin{align}
\label{6.McM}{{M}_{c}}=&\left(1+R\right)^2,\\
\label{6.M0}{{M}_{0}}(\tau)=&2R  \exp \left[ -\frac{{{\left( T-\tau  \right)}^{2}}}{2{{\Delta }^{-2}}} \right] \cos\left[2 \Omega(T-\tau)\right],\\ \nonumber
{{M}_{1}}(\tau)=&-\frac{2\sqrt{2}r(1+R)}{\Delta\sqrt{\kappa}}\left(
\exp \left[ -\frac{{{\left( T-\tau +2 \kappa \Omega \right)}^{2}}}{2{{\kappa }^{2}}{{\Delta }^{2}}}
\right]
+
\exp \left[ -\frac{{{\left( T-\tau -2 \kappa \Omega \right)}^{2}}}{2{{\kappa }^{2}}{{\Delta }^{2}}}
\right]\right)\times\\ \label{6.M1}
&\cos \left[{{\omega }_{0}}\left( T-\tau  \right)- \frac{{{\left( T-\tau  \right)}^{2}}}{4\kappa }+\frac{\pi}{4} \right],\\
\label{6.M2}{{M}_{2}}(\tau)=&\frac{2R^2}{\Delta \sqrt{\kappa }}
\exp\left[ -\frac{2 \Omega^2}{\Delta^2} \right]
\exp \left[ -\frac{{{\left( T-\tau  \right)}^{2}}}{2{{\delta }^{-2}}} \right]
\cos \left[2{{\omega }_{0}}\left( T-\tau  \right)+\frac{\pi}{4} \right].
\end{align}
\end{subequations}

The total interferogram, its terms, and their spectra are plotted in Fig.~\ref{fig:TheorPlots}d. Again, the spectra $\abs{\Tilde{M}(\omega)}$ look similar to the case without dispersion (Fig.~\ref{fig:TheorPlots}b). 
Corresponding interferogram terms $M_0$ and $M_2$ also resemble the structure shown in Fig.~\ref{fig:TheorPlots}a. However, the amplitude of the $M_2$ term is $\Delta\sqrt{\kappa}\exp{2\Omega^2/\Delta^2}$ times smaller. The $M_1$ term is again not Fourier-limited. Its envelope exhibits two peaks located at $\pm 2\kappa\Omega$ with an equal width $2\kappa\Delta$. Therefore, even for the non-degenerate case, one can separate one can separate the $M_0$ term and achieve dispersion cancellation under the condition $2\Omega+\Delta\ll \omega_{\text{p}}/3$.

\section{Experimental Data Processing}

\subsection{Interferogram processing}
\begin{figure}[ht]
    \centering
    \includegraphics[width= 1\textwidth]{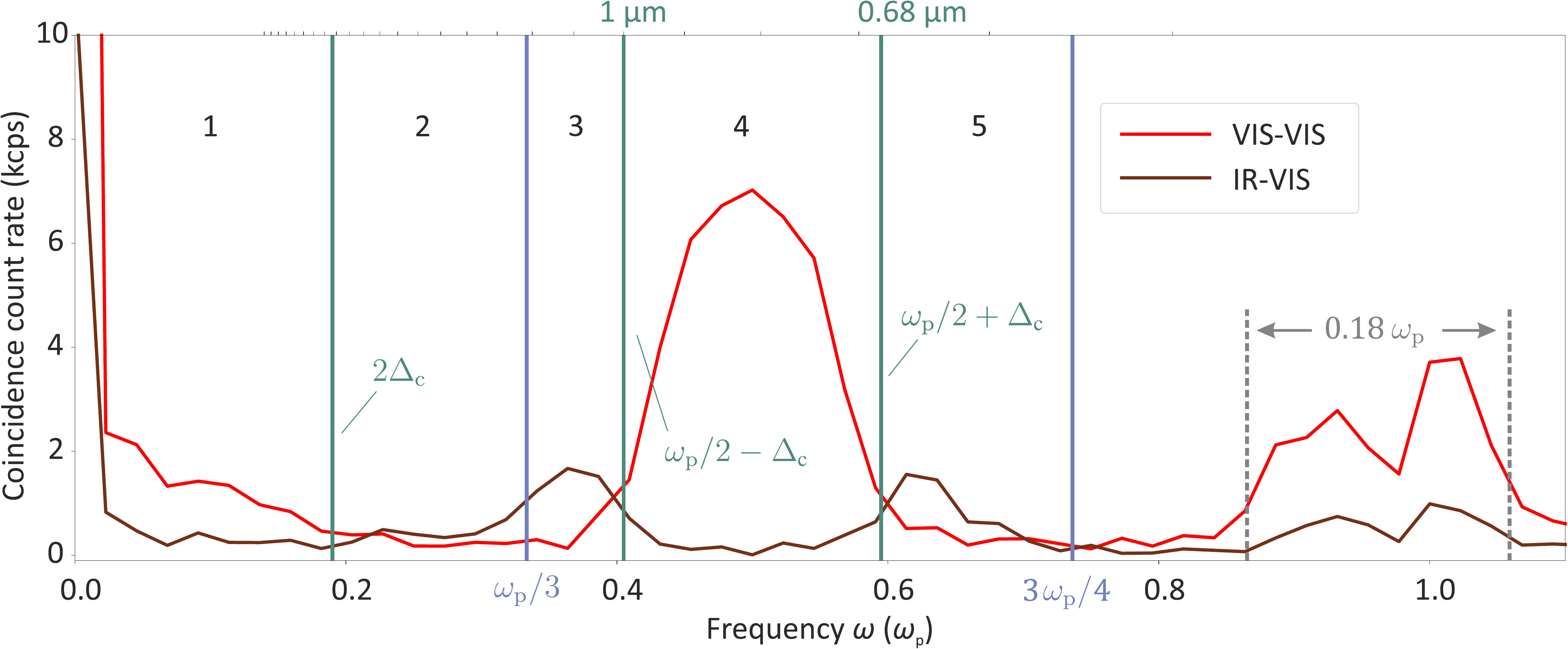}
    \caption{ Fast-Fourier transform of raw interferograms.}
    \label{fig:fft}
\end{figure}

To process the coincidence interferograms without dispersion, presented in Fig.~4b of the main text, to select the $M_0$ and $M_1$ terms and to account for detection efficiencies, we calculated their Fast-Fourier Transform (FFT). The absolute values of the obtained spectra are shown in Fig.~\ref{fig:fft} and in Fig.~4c of the main text.

The red curve (VIS-VIS) represents the coincidences between two visible detectors (Laser Components COUNT NIR) and corresponds to the degenerate case (Fig.~\ref{fig:TheorPlots}a), while the brown curve (IR-VIS) represents the coincidences between visible and infrared  (MPD PDM-IR) detectors and corresponds to the non-degenerate case (Fig.~\ref{fig:TheorPlots}b). 

Both the VIS-VIS and IR-VIS spectra exhibit a peak at $\omega_\text{p}$, corresponding to the $\Tilde{M}_2$ term.

The VIS-VIS spectrum exhibits a broadband peak centered at $\omega_\text{p}/2$, corresponding to the $\Tilde{M}_1$ term, equivalent to single-photon interference in a Michelson Interferometer (MI).
In the case of IR-VIS coincidence detection, which is equivalent to selecting a non-degenerate part of the biphoton spectrum, the $\Tilde{M}_1$ term is represented by two peaks (consistent with Fig.~\ref{fig:TheorPlots}b). 
These peaks are separated from the VIS-VIS $\Tilde{M}_1$ peak by the frequency $\omega_\text{c} \equiv \omega_\text{p}/2 - \Delta_\text{c}$, corresponding to the dichroic mirror cut-off wavelength $\lambda_\text{c} = 1\ \rm\mu m$ and its conjugate frequency $\omega_\text{p}/2 + \Delta_\text{c}$.

Finally, the VIS-VIS interferogram exhibits a peak centered at zero frequency, corresponding to the Hong-Ou-Mandel (HOM) peak, $\Tilde{M}_0$. 
The same term is represented in the IR-VIS interferogram as a shifted peak (also in agreement with Fig.~\ref{fig:TheorPlots}b), separated from the VIS-VIS $\Tilde{M}_0$ peak by the frequency $2\Delta_\text{c}$.

\begin{figure}[ht]
    \centering
    \includegraphics[width= 0.8\textwidth]{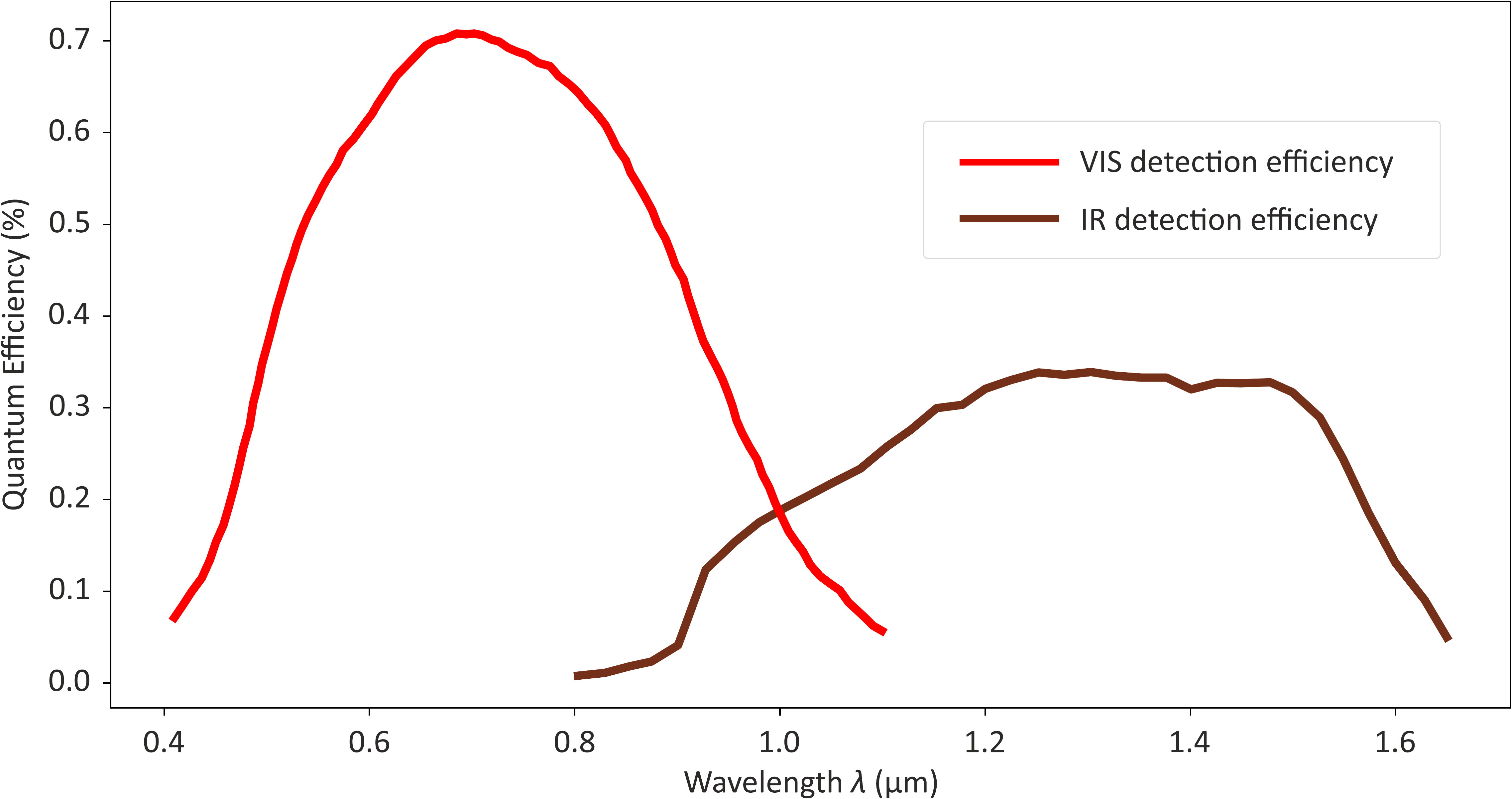}
    \caption{Detection efficiency of Laser Components COUNT NIR (red curve) and MPD PDM-IR (brown curve).}
    \label{fig:efficiency}
\end{figure}

To account for detection efficiency, we use the data provided by the manufacturers of our detectors (Laser Components COUNT NIR and MPD PDM-IR), shown in Fig.~\ref{fig:efficiency}. 

To separate different spectral terms and isolate the impacts from different detectors, we divide the entire spectral range into five zones, presented in Table~\ref{tab:coincidences_efficiency}. Since the terms $\Tilde{M}_0$ (zones 1, 2) and $\Tilde{M}_1$ (zones 3--5) are partially overlapped, we separate them by the frequency $\omega_\text{p}/3$. The right limit of zone~5, $\omega_\text{p}-\omega_\text{lim}$, is defined by the right limit of the available data for the IR detection efficiency, $\lambda_\text{lim} = 1650$~nm. 

\begin{table}[b]
 \caption{Coincidences efficiency calibration.\label{tab:coincidences_efficiency}}
\begin{tabular}{|c|c|c|c|r|}
\hline
Zone& Term & Frequency range & Data taken for correction &
  Efficiency correction                                          \\ \hline

1&$\Tilde{M}_0$  & $0<\omega<2\Delta_\text{c}$  & $\vphantom{\int\limits_0^2}$ VIS-VIS   
& $\frac{1}{2}\times\eta_\text{VIS}\left( \frac{\omega_\text{p}+\omega}{2}\right) \eta_\text{VIS}\left( \frac{\omega_\text{p}-\omega}{2}\right)$ \\ \hline

2&$\Tilde{M}_0$  & $2\Delta_\text{c}<\omega<\frac{\omega_\text{p}}{3}$ & $\vphantom{\int\limits_0^2}$ IR-VIS    &$\eta_\text{VIS}\left( \frac{\omega_\text{p}+\omega}{2}\right) \eta_\text{IR}\left( \frac{\omega_\text{p}-\omega}{2}\right)$ \\ \hline

3&$\Tilde{M}_1$  & $\frac{\omega_\text{p}}{3} < \omega < \frac{\omega_\text{p}}{2}-\Delta_\text{c}$ & $\vphantom{\int\limits_0^2}$ IR-VIS    & $\eta_\text{IR}(\omega)\ \eta_\text{VIS}(\omega_\text{p}-\omega)$      \\ \hline

4&$\Tilde{M}_1$  & $\frac{\omega_\text{p}}{2}-\Delta_\text{c}<\omega<\frac{\omega_\text{p}}{2}+\Delta_\text{c}$ & $\vphantom{\int\limits_0^2}$ VIS-VIS   & $\frac{1}{2}\times\eta_\text{VIS}(\omega)\ \eta_\text{VIS}(\omega_\text{p}-\omega)$   \\ \hline

5&$\Tilde{M}_1$  & $\frac{\omega_\text{p}}{2}+\Delta_\text{c}<\omega<\frac{3\omega_\text{p}}{4}$ &  $\vphantom{\int\limits_0^2}$ IR-VIS    & $\eta_\text{VIS}(\omega)\ \eta_\text{IR}(\omega_\text{p}-\omega)$    \\ \hline
\end{tabular}
\end{table}

The data in each frequency zone are normalized by the efficiency correction terms listed in the fifth column of Table~\ref{tab:coincidences_efficiency}. Note that the correction terms for VIS-VIS coincidences have an additional factor of $1/2$, since in half of all cases, two visible photons go to the same detector and contribute nothing to the coincidences.

The processed $\tilde{M}_1$ term, presented in Fig.~4c with a bold dotted curve, combines the normalized VIS-VIS spectrum from zone~4 and the normalized IR-VIS spectra from zones~3 and 5. Its measured bandwidth was estimated as $\text{FWHM}_\omega = 2\pi \times 136~\text{THz}$.

However, this spectral peak may appear broader than expected due to phase fluctuations and piezo stage inaccuracies in the experimental setup. To estimate the extent of this broadening effect, consider the $\tilde{M}_2$ peak at $\omega_\text{p}$. According to theory, this peak should be as narrow as the pump laser linewidth, but the experimentally observed $\tilde{M}_2$ peak exhibits a significant bandwidth of $\text{FWHM}_\text{p} = 0.18 \times \omega_\text{p}$. This corresponds to a wavelength bandwidth of $\text{FWHM}_\lambda = 0.18 \lambda_\text{p} = 73~\text{nm}$, which reflects the typical fluctuations in the optical path length of the interferometer arms.

At a frequency of $\omega_\text{p}/2$, this wavelength broadening translates into a frequency broadening of:
\begin{equation}
  \text{FWHM}_\text{broad} =  \frac{\omega_\text{p}}{2} \times \frac{\text{FWHM}_\lambda}{2 \lambda_\text{p}} = \frac{\text{FWHM}_\text{p}}{4}.  
\end{equation}
Taking this into account, the corrected bandwidth of the $\tilde{M}_1$ term can be calculated as:
\begin{equation}
\text{FWHM}_\text{corrected} = \sqrt{\text{FWHM}^2 - \text{FWHM}_\text{broad}^2} = 2\pi \times 132~\text{THz}.
\end{equation}
This correction suggests that the experimentally measured spectral peak is broadened due to systematic factors, such as mechanical inaccuracies and phase fluctuations, and the corrected value is consistent with theoretical expectations.



The processed $\tilde{M}_0$ term, presented in Fig.~4c with a bold solid curve, combines the normalized VIS-VIS spectrum from zone~1 and the normalized IR-VIS spectrum from zone~2. Its inverse FFT gives the processed interferogram, shown in Fig.~4b as a blue curve.

The coincidence interferograms with dispersion, presented in Fig.~4e, were processed in the same way.

For single counts, we also calculate FFT spectra of raw interferograms, divide the data for the visible and IR detectors by their corresponding efficiencies, combine the data from the IR detector for $\omega<\omega_\text{c}$ with the data from the VIS detector for $\omega>\omega_\text{c}$, and then calculate the inverse FFT to obtain a corrected interferogram, presented by blue curve in Fig.~4a,d.



\subsection{Generation efficiency estimation}

The coincidence count rate measured with two visible detectors before the entrance of the interferometer (without any filters) was $R_\text{detected}=3$~kcps for a pump power of $P=8$~mW.

The losses caused by the VIS detectors' quantum efficiency can be estimated as follows. Consider the processed data for the MI-term $\Tilde{M}_1$, combined from both the VIS-VIS and IR-VIS normalized spectra, shown in Fig.~4c with a bold dotted curve. Its integral over $\omega$ is proportional to the generated coincidence rate $R_\text{generated}$. To account for the VIS detector efficiency, we multiply $\Tilde{M}_1$ by the normalization factor presented in Table~\ref{tab:coincidences_efficiency}, and its integral over $\omega$ is proportional to the detected coincidence rate $R_\text{detected}$.

Therefore, the source efficiency is estimated as:
\begin{equation}
    \frac{R_\text{generated}}{P}=\frac{2 \int \Tilde{M}_1(\omega)\text{d}\omega}{\int \Tilde{M}_1(\omega) \eta_\text{VIS}(\omega)\ \eta_\text{VIS}(\omega_\text{p}-\omega)\text{d}\omega}\frac{R_\text{detected}}{P} \approx  2.7 \ \frac{\text{ kcps}}{\text{mW}}.
\end{equation}

To estimate the source's spectral coincidence efficiency $S_0/P$, we divide the obtained value of the source efficiency ${R_\text{generated}}/{P}$ by the integral bandwidth $B_\omega = \sqrt{\frac{\pi}{4 \ln 2}}\text{FWHM}_\omega = 2\pi\times 141$~THz:
\begin{equation}\label{eq:spectral_coinc_countrate}
    \frac{S_0}{P}= \frac{R_\text{generated}}{B_\omega P }\approx  20 \ \frac{\text{ cps}}{\text{THz} \times \text{mW}}.
\end{equation}

Here and throughout the text, we use the following relations between different bandwidth measures for a Gaussian peak:
\begin{subequations}
\begin{align}
    \text{Gaussian function}  & = \exp \left(-\frac{x^2}{2 \Delta^2}\right),\\
    \text{Standard deviation (STD)} & = \Delta,\\
    \text{Full width half maximum (FWHM)} & = \sqrt{8\ln 2} \Delta = \sqrt{\frac{4 \ln 2}{\pi}}\ B, \\
    \text{Integral bandwidth}\ B& = \sqrt{2\pi} \Delta = \sqrt{\frac{\pi}{4 \ln 2}}\ \text{FWHM}.
\end{align}
\end{subequations}




